\documentclass[namedreferences]{solarphysics}
\usepackage[hyperref,optionalrh,showbiblabels]{spr-sola-addons} 
\usepackage{sidecap}
\usepackage{graphicx}        
\usepackage{color}           
\usepackage[shortlabels]{enumitem}

\chardef\us=`\_

\begin{document}

\begin{article}
\begin{opening}

\title{Image Quality Specification for Solar Telescopes }
\author[addressref={aff1},corref,email={saraswathi.kalyani@iiap.res.in}]{\inits{S. K.}\fnm{Saraswathi Kalyani}~\lnm{Subramanian}\orcid{https://orcid.org/
0000-0002-2508-3653}}
\author[addressref=aff1,email={sridharan.r@iiap.res.in}]{\inits{S.}\fnm{Sridharan}~\lnm{Rengaswamy}\orcid{https://orcid.org/0000-0001-7050-7518}}

\address[id=aff1]{Indian Institute of Astrophysics}

\runningauthor{S. K. Subramanian and S. Rengaswamy}
\runningtitle{Image Quality Specification}

\begin{abstract}

Modern large ground-based solar telescopes are invariably equipped with adaptive optics systems to enhance the high angular resolution imaging and spectroscopic capabilities in the presence of the Earth's atmospheric turbulence. The quality of the images obtained from these telescopes can not be quantified with the Strehl ratio or other metrics that are used for nighttime astronomical telescopes directly. In this paper, we propose to use the root mean square (rms) granulation contrast as a metric to quantify the image quality of ground-based solar telescopes.  We obtain semi-logarithmic plots indicating  the correspondence between the Strehl ratio and the rms granulation contrast for most practical values of the telescope diameters ($D$) and the atmospheric coherence diameters ($r_0$), for various levels of adaptive optics compensation.   We estimate the efficiency of a few working solar adaptive optics systems by comparing the results of our simulations with the Strehl ratio and rms granulation contrast published by these systems. Our results can be used in conjunction with a plausible 50\% system efficiency to predict the lower bound on the rms granulation contrast expected from ground-based solar telescopes.
\end{abstract}
\end{opening}

\section{Introduction}
     \label{S-Introduction} 
The image quality of ground-based solar telescopes has been studied in detail over several decades starting from the seminal work of \cite{1964IAUS...19..193K}. With the advent of the theory on the effects of the atmospheric turbulence on the ground-based telescopes (\cite{Roddier1981} and the references therein) it is now well understood that the image quality is characterised with a single parameter known as the Fried's parameter ($r_0$) or the atmospheric coherence diameter. In ground-based solar imaging, $r_0$ is directly linked with the contrast of the images \citep{Roddier1981}. It is interesting to note that the use of image contrast as a measure of image quality was proposed by Kiepenheuer(\citeyear{1964IAUS...19..193K}) even before the invention of the parameter $r_0$.

Modern large solar telescopes \citep{2003SPIE.4853..370S, BBSO, GREGOR_AO, Fuxian_Obs, MAST, DKIST} are invariably equipped with adaptive optics (AO) systems to mitigate the deleterious effects of the Earth's atmospheric turbulence on the image quality.  Understandably, the performance of the adaptive optics systems depend on the characteristic parameters of the atmosphere, namely, $r_0$ and the atmospheric coherence time $\tau_0$.

One of the parameters used for characterising the performance of an AO system is the Strehl ratio. It is the ratio of the on-axis intensity of an aberrated point spread function (PSF; here the term aberrated actually means the residual aberrations after AO corrections)  to the on-axis intensity of a diffraction-limited PSF. It has a maximum value of unity and a high value implies better performance of the system. 

For stellar telescopes, since most objects are point sources, this parameter is useful in characterising the associated AO systems. However, in the case of solar telescopes where the object is extended in nature, this parameter cannot be determined easily. 

Under such circumstances, how can one possibly quantify the performance of a large ground-based solar telescope? What is the metric that could be specified as a requirement for the telescope (for example, to a telescope manufacturer/vendor)?
As AO is likely to be working efficiently only under good atmospheric conditions (large $r_0$ and high $\tau_0$), what is the metric that can be used for image quality: (a) when conditions are not optimal for AO to be operational (no-AO mode), (b) when conditions are optimal for AO  but only a partial compensation of the wavefront distortion is achieved (low order AO), (c) when a complete compensation is achieved  under good atmospheric conditions?   Answering these questions in as much quantitative nature as possible is the main motivation behind this work.

There are many parameters that are used to characterise the image quality of solar telescopes \citep{2017SoPh..292..187P}. The image quality metrics (IQM) are broadly classified into three categories - full reference, reduced reference and no-reference methods \citep{2015SoPh..290.1479D}. As the names suggest, the first two metrics are completely or partially dependant on a reference image based on which the quality of other images can be determined. The metrics of the last category are completely independent of a reference image. In the case of ground-based telescopes, where the image is corrupted by the atmosphere and instrumental effects, it is difficult to select an ideal reference image. Therefore, no-reference IQM are preferred. One such metric is the Median Filter Gradient Similarity (MFGS) proposed by \cite{2015SoPh..290.1479D}. In this method, the gradients of the instantaneous images whose quality are to be determined and that of their filtered versions are compared to determine the MFGS parameter. It varies between zero and unity and it was shown that a larger value corresponds to better seeing conditions. However, we found that MFGS works only for short exposure images (instantaneous images) and is therefore useful to sort and then select the best images from an observation (for example, \cite{2018SoPh..293...44D}), which can then be post-processed using techniques like speckle masking to further improve the image quality. While MFGS is useful for such an application and is also superior to rms granulation contrast due to its relative independence on the scene, it cannot be used for our application as we are determining the rms granulation contrast of solar images which have exposure times several times larger than the daytime atmospheric coherence time. The assumption we have made is that without AO, the exposure time is short enough not to be affected by telescope errors and with AO the errors will be corrected by the tip-tilt system. In other words, we wanted to arrive at a metric equivalent to that of the Strehl ratio which is defined only for long exposure images. Therefore, we also look for a metric which will work for long exposure images.

Therefore, in this paper, we use the rms granulation contrast as a metric to specify the image quality both in the presence and absence of the adaptive optics compensation. Although there is quite a bit of intrinsic variation of the rms granulation contrast \citep{SST}, its choice as a metric is quite useful and practical owing to the presence of granulation throughout the solar disk independent of the solar activity cycle. The intrinsic low contrast of the granulation prevent their use for adaptive optics wavefront sensing when their observed contrast is further lowered due to poor observing conditions (low $r_0$). 

The reminder of this paper is organised as follows. In Section~\ref{sec-sim}, we start with the formal definition of the rms granulation contrast and describe the details of our simulation. In Section~\ref{Sec:ResAndDis} we present and discuss the results. In Section \ref{summary} we present a summary.

\section{Simulations and Validation}
\label{sec-sim}

\subsection{The metric}

We use the rms contrast ($C_{rms}$) of the solar granulation as a metric. The metric is always applied to an average long exposure image (with or without AO correction, as the case maybe) normalised to unit mean intensity (measured in counts). For a two dimensional digital image array $s(x,y)$, it is defined as
\begin{equation}
C_{rms} = \left[\frac{s-\langle s \rangle}{\langle s \rangle}\right]_{rms},
\end{equation}
where $\langle~\rangle$ indicates spatial average and $s(x,y)$ could be an object intensity distribution or an image intensity distribution.

\subsection{The Object Model}

We used simulated solar granulation image against space or ground-based image as input object in our simulations. This choice was driven by the fact that neither space-based images nor ground-based images\textemdash{AO corrected alone or AO plus speckle or phase-diversity corrected}\textemdash are likely to be completely free from residual instrumental effects and thus may bias the results based on their use.  Further, with the advent of advanced cluster computational facilities, we are now able to synthesis instantaneous bolometric solar granulation images fairly accurately. The high-resolution structure of the solar photospheric granulation employed in this work is a snapshot of the bolometric intensity from a 3D numerical simulation carried out with the radiative magnetohydrodynamic (MHD) code CO5BOLD \citep{2012JCoPh.231..919F}, which solves the coupled system of compressible MHD equations that include an imposed gravitational field and non-local, frequency-dependent radiative transfer. The simulations were performed on a 3D Cartesian box of size 9.6 x 9.6 x 2.8 Mm$^3$, with a uniform grid size of 15 x 15 x 10 km$^3$ in (x,y,z). The vertical domain ranges from about 1300 km below the optical depth $\tau_{500}$ = 1 surface (photosphere) to 1500 km above it in the chromosphere and the gravitational field is uniform and vertical with values of log (g) = 4.44. A constant entropy inflow is supplied at the bottom boundary of the simulation domain to maintain an average surface effective temperature of $T_{eff}$ = 5770 K. The simulation box set up as above was derived from a CO5BOLD simulation performed by \cite{2016A&A...596A..43C}, and the computations were carried out by S.P. Rajaguru (private communication) using IIA's HPC (Indian Institute of Astrophysics' High Performance Computing) cluster Nova.  

It should be emphasized that we use only a small segment of the  high resolution solar granulation image for the work presented in this article and the segment size, which is basically the field-of-view, is set by the other simulation parameters (See Section~\ref{subsec:fov}).

\subsection{The Atmospheric Model}

We model the phase perturbations induced by the atmosphere through a two dimensional phase-screen generated using the Kolmogorov model of turbulence \citep{2004SPIE.5171..219S,2010SoPh..266..195P,Merak}. The phase-screen is characterised by the Fried's parameter $r_0$. A large phase-screen of size 163.84 $\times$ 163.84 m$^2$ is simulated with a pixel sampling of $2$ cm. Assuming frozen-field approximation, the phase-screen is blown past the telescope aperture and several thousands of segments of the phase-screen with size equal to the size of the aperture are used.

In our simulations, we generated phase screens with $r_0$ varying discretely between 6 cm and 21 cm (at $\lambda$ = 430.5 nm) with a step size of 1 cm, to account for the wide range of the daytime seeing conditions.  
\subsection{The Instrument Model}
\subsubsection{The Telescope}
We used an un-obscured 2 dimensional pupil function $W({\bf x})$ described by
\begin{equation}
       W({\bf x})=\left\{
                \begin{array}{ll}
                  1, \vert {\bf x}\vert  \le D/2\\
                  0, \vert {\bf x}\vert > D/2
                \end{array}
              \right.
\end{equation}
to model the telescope, where ${\bf x}$ represents the spatial coordinates at the pupil and $D$ is the aperture size.  Several distinct values of pupil diameters were used starting from 30 cm to 200 cm, the values were chosen to represent the aperture size of the solar telescopes available in India and elsewhere in the world.

\subsubsection{Modelling AO corrections}
\label{AO_Corr}

We implemented the AO correction in an idealistic way. We decomposed the instantaneous phase-perturbations over the pupil into a given number, $N_Z$, of Zernike Polynomials \citep{1976JOSA...66..207N} using a least-square solution method \citep{svd}. A model phase front (phase perturbations over the pupil) was then synthesized from the Zernike coefficients obtained from the least square method and subtracted from the initial perturbed phase-front to obtain the residual phase perturbations after ``AO correction". As we sampled the phase screen with a sampling of 2 cm, and made a pixel-wise correction, it is an ideal correction up to 2 cm. Perturbations on spatial scales less than 2~cm are neither generated nor compensated in our approach. The number of Zernike polynomials used to model a given phase perturbation increased in steps of two radial orders at a time (2, 9, 20, 35, and so on) until the variance of the residual phase perturbations over the pupil becomes less than 1 radian$^2$. This criterion essentially enables us to terminate the simulation at a particular $r_0^{\prime}$ (that could be any value between 6 and 21) for a given $D$.

\subsection{Simulation Flow}
\label{simulation_flow}

We assume unit amplitude perturbations and obtain the instantaneous PSF as the modulus squared Fourier transform of the pupil-plane phase distribution (product of the ideal pupil function $W({\bf x})$ and the phase perturbations $\phi({\bf x}$) represented by a single segment of the phase screen of a given $r_0$) expressed in complex exponential form $W({\bf x})\exp\left[j\,\phi({\bf x})\,W({\bf x})\right]$.
 We obtain the instantaneous transfer function as the inverse Fourier transform of the area normalized instantaneous PSF. We model the instantaneous image as the convolution of the object intensity distribution with the instantaneous PSF. In practice, the convolution is achieved through an inverse Fourier transform of the product of the Fourier transform of the object intensity distribution and the instantaneous transfer function.
 
The instantaneous images are averaged over 1000 realizations of the atmospheric phase perturbations and an average image is obtained. The process is repeated 10 times so that we can obtain an average rms contrast and the variations associated with it. We found the variations in rms contrast to be much smaller than a significant fraction of the actual contrast. So, they were not visible when plotted as error bars. The process is repeated for AO corrected images as well to get the corresponding values. Finally, we added the photon noise to the images. However, we found that rms granulation contrast is immune to this photon noise.
 
\begin{figure}[H]  

               \includegraphics[trim = {0 2.75cm 0 2.75cm}, width=0.7\textwidth,clip=]{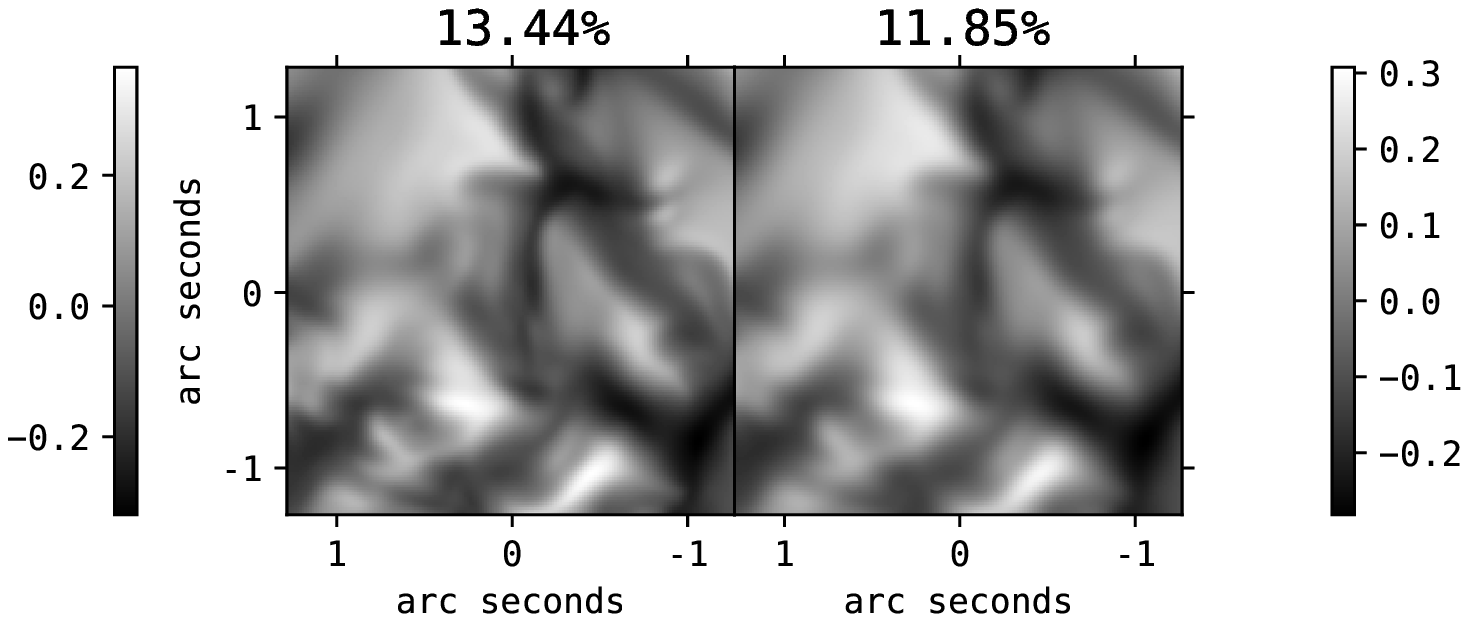}

               \includegraphics[trim = {0 2.75cm 0 3cm}, width=0.7\textwidth,clip=]{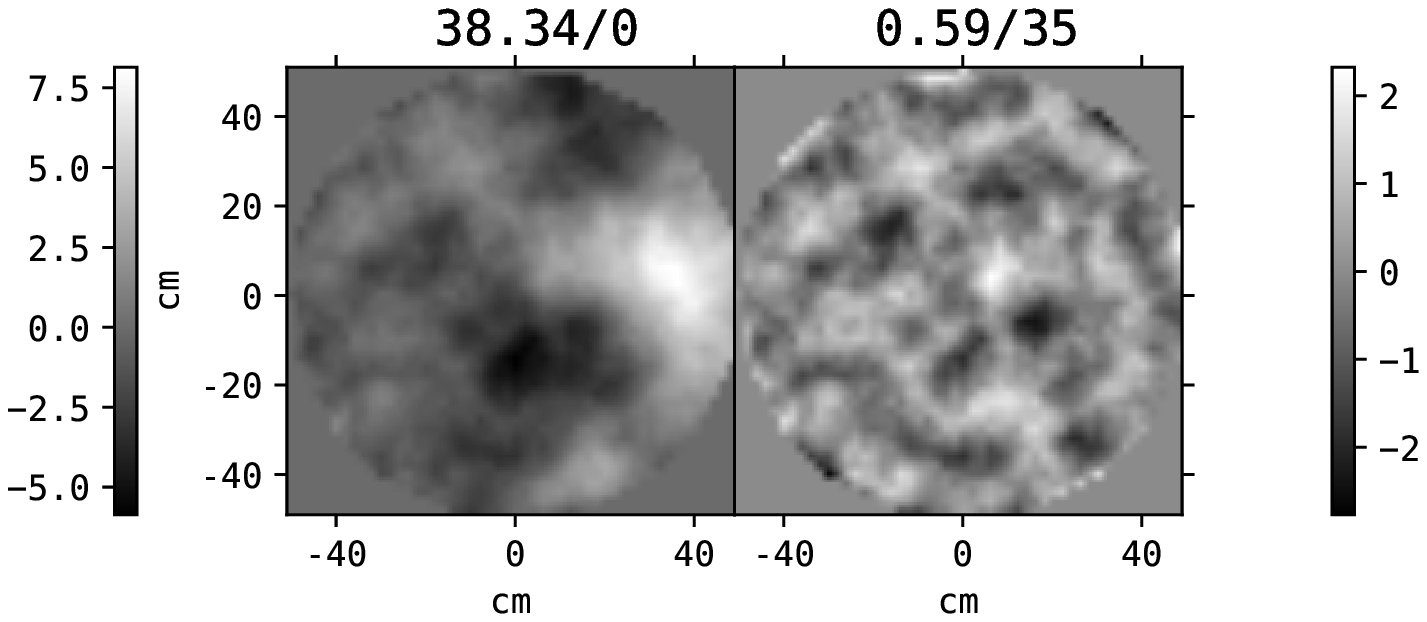}

                \includegraphics[trim = {0 0 0 0.7cm}, width=0.7\textwidth,clip=]{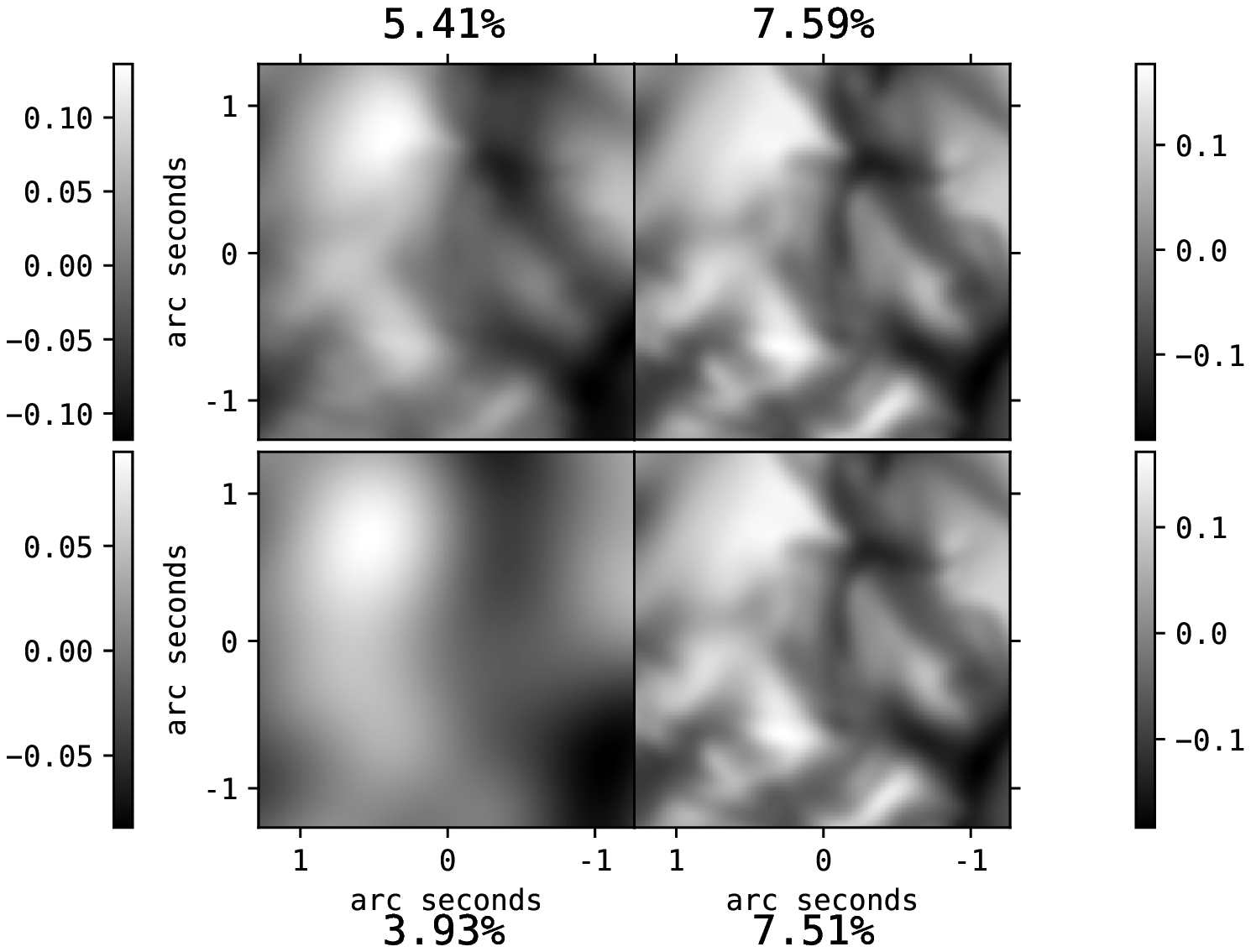}

\caption{Top panels indicate the object (left) and the diffraction-limited (right) image intensity distributions. Panels on the second row indicate the residual phase distributions without (left) and with (right) AO correction.  The mean square residual phase variance/number of Zernike terms for which the correction is done are indicated on the top. Panels on the third row indicate instantaneous images without (left) and with AO correction (right). Panels on the last row show the long exposure images without (left) and with AO correction (right). The quantities expressed in percentage are the rms contrast values for the respective images.}
   \label{composite}
   \end{figure}

Thus, the main free or input parameters in our simulation are: the telescope diameter $D$, the atmospheric coherence diameter $r_0$, the number of compensated Zernike terms $N_Z$.  The metric we use to characterise the average image (with and without AO correction as the case may be) is the rms granulation contrast $C_{rms}$. In addition, we also use the traditional metrics like the residual mean square phase variance over the pupil after AO correction and the Strehl ratio of an average stellar PSF.
Figure~\ref{composite} shows the flow of the simulation. The top left and right panels indicate the object (simulated solar granulation) and the diffraction-limited image intensity distributions, respectively. The second row of panels indicate the pupil-plane phase distributions. The images on the bottom panels indicate instantaneous and long exposure images (third and fourth rows) respectively without (left) and with (right) AO correction.

\subsubsection{Field-of-View and Wavelength Dependency}
\label{subsec:fov}

As we perform the Fourier transform using Fast Fourier Transform routines that keep the number of pixels same in either domain,  we constrain the simulation window size to be at least twice that of the aperture size, with the aperture centered on the window, so that the transfer function is not truncated. As our simulations involved different aperture sizes, we choose a window size of 256 $\times$ 256 pixels so that apertures of up to 128 pixels could be simulated. 
 This leaves us with the feasibility of performing simulations up to 2.56 m diameter aperture, for a 2 cm pixel sampling.  However, we have performed simulations only up to $D = $ 2 m. 
 As a result of the Fourier transformation relation between the pupil plane and the image plane, the 
field-of-view (in the image plane), determined by the pupil-plane pixel sampling, is $[\lambda/0.02]^2 \approx 4.4^{\prime\prime} \times 4.4^{\prime\prime}$ arcsec$^2$ at 430.5 nm. As the atmospherically induced pupil-plane phase perturbations are independent of the wavelength, $\lambda$ becomes an independent parameter in our simulations. However, for the purpose of choosing the field-of-view, we have chosen the wavelength to be 430.5 nm. This also means that the $r_0$ values used in our simulations correspond to 430.5 nm.

\subsection{Validation}

\begin{figure}[H]    

    \includegraphics[trim = {0 0 1cm 0},
              width=0.7\textwidth,clip=]{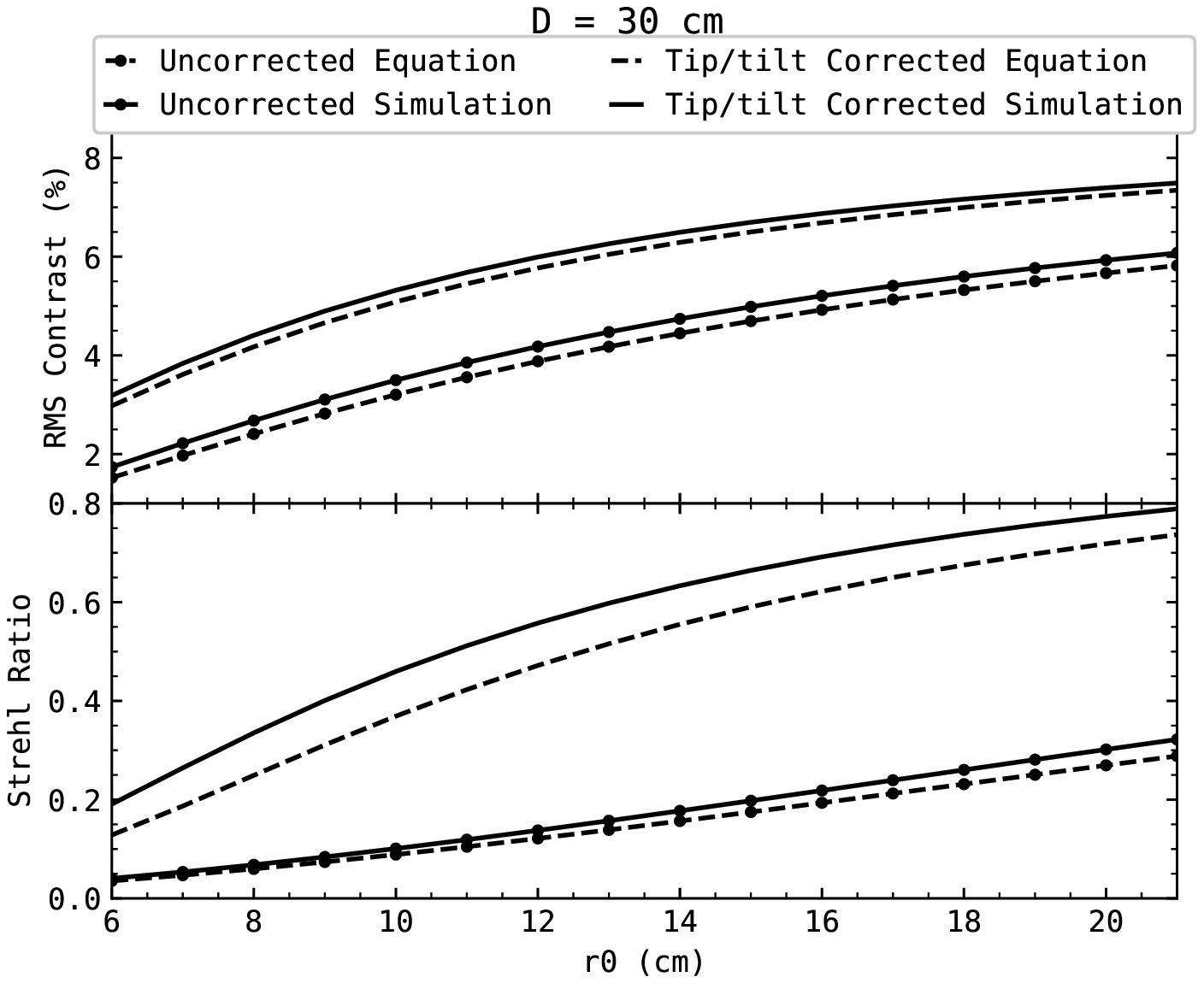}

    \includegraphics[trim = {0 0 1cm 0}, width=0.7\textwidth,clip=]{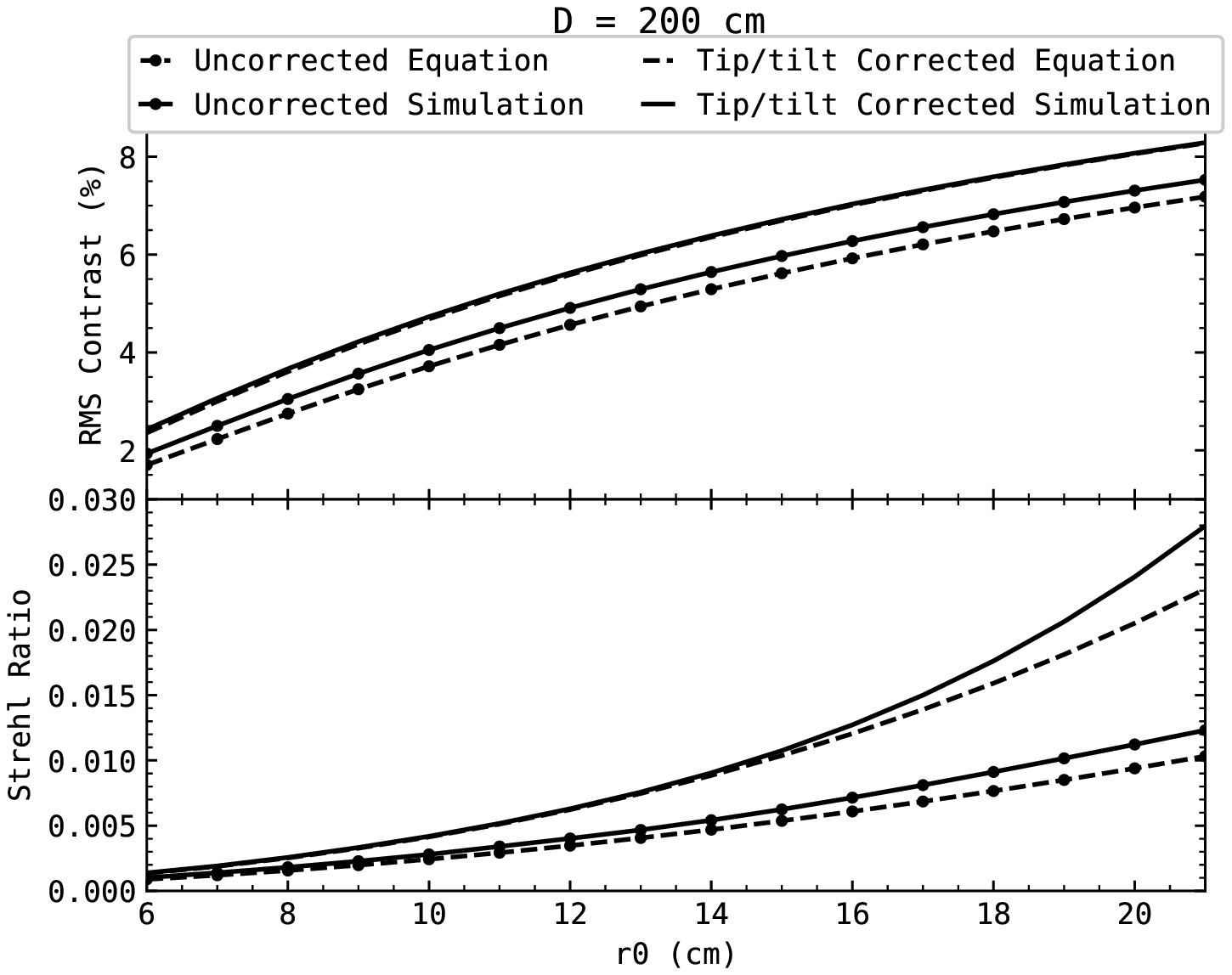}

\caption{Rms Contrast and Strehl ratio as a function of $r_0$ for two values of the diameter (30 and 200 cm). The circular marked and unmarked curves are for the uncorrected (seeing-limited) and tip-tilt corrected (image stabilized) values respectively. The values obtained from the analytical expressions of the transfer function are shown by the dashed curves and those from the simulation are shown by the solid curves.}
  \label{equ_and_sim}
  \end{figure}

We checked the veracity of our simulation procedure and the results for cases in which they could be obtained with analytical expressions,  in two distinct ways. First, from theory, we know the exact analytical expressions for average long and tilt-corrected  short exposures transfer functions \citep{1966JOSA...56.1372F,Roddier1981}. We could reproduce these functions through our simulations. Further, we estimated the Strehl ratio as the ratio of the volume under the transfer functions to that of the ideal diffraction-limited transfer functions both for uncorrected (seeing-limited) and tip/tilt corrected cases.   We know that for a large telescope ($D \gg r_0$), under seeing-limited imaging conditions, the Strehl ratio can be approximated as $(r_0/D)^2$. We found that our simulated Strehl values  were in good match with the theoretical values. The bottom panels of Figure~\ref{equ_and_sim} show the Strehl ratios derived from the analytical expressions for the transfer functions and the simulations for two telescope diameter values.
We see that: (a) the values are close to what is to be expected, (b) the Strehl values derived from the simulation are always slightly higher than that predicted by the theory. We already know that the theoretical expressions for average short exposure transfer functions are overestimated at high spatial frequencies as they are derived under the assumption that there is no correlation between the tilt and high order wavefront perturbations \citep{tip_tilt_paper,1977ApOpt..16..665L}.

The top panels of Figure~\ref{equ_and_sim} show the rms granulation contrast estimated as a function of $r_0$ with the simulated long and short exposure transfer functions for two representative cases of diameters (30 cm and 200 cm). The corresponding values estimated using the analytical expressions of long and short exposure transfer functions are overplotted.  We find that the rms contrast estimated from our simulations closely follow that expected from the analytical expressions of the transfer functions.

Further, we decomposed the atmospheric phase perturbations over the pupil into a finite number of Zernike polynomials and found that the residual phase variance after compensating for a certain number of Zernike terms is always slightly less than the corresponding theoretical value predicted by \cite{1976JOSA...66..207N}. The maximum difference is less than 10\%.

This validates our simulation procedure and allows us to estimate the rms contrast for high order phase compensated images where there are no analytical expressions available.

\section{Results and Discussion}
\label{Sec:ResAndDis}

\subsection{Results}

\subsubsection{Seeing-limited Imaging}
   \label{Seeing_ltd} 
\begin{SCfigure}[][h]  
    \caption{Rms contrast vs Fried's parameter $r_0$  under seeing-limited imaging conditions where the markers represent the values of diameter (in cm) as shown in the legend}.
    \includegraphics[width=0.6\textwidth]{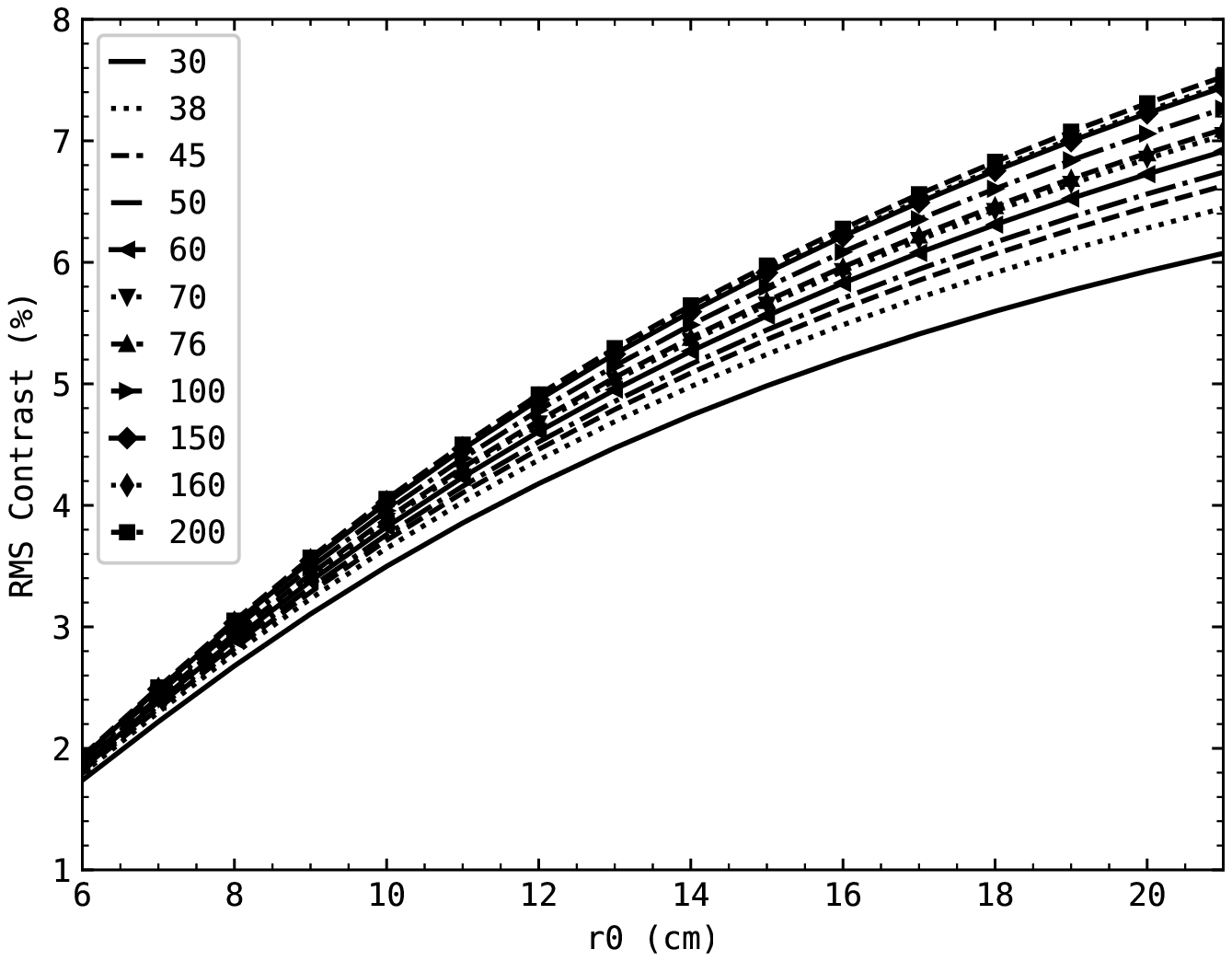}
\label{seeing_ltd}
\end{SCfigure}

   Figure~\ref{seeing_ltd} shows the rms contrast as a function of $r_0$ for different telescope sizes under seeing-limited imaging. We find that it varies between 1.8 and 7.6\%. It increases with $r_0$ as expected. It has a slight dependence on the telescope diameter as well; larger the diameter, higher the rms contrast, for a given $r_0$.
   
\begin{figure}[H]  
    \caption{Rms contrast vs Strehl ratio under seeing-limited imaging conditions. The markers correspond to different telescope diameters (in cm) as shown in the legend and the direction of increasing $r_0$ is along the arrow ranging from 6 to 21 cm in steps of unity. }  
    \includegraphics[trim = {0 0.5cm 1.65cm 0}, width=0.55\textwidth]{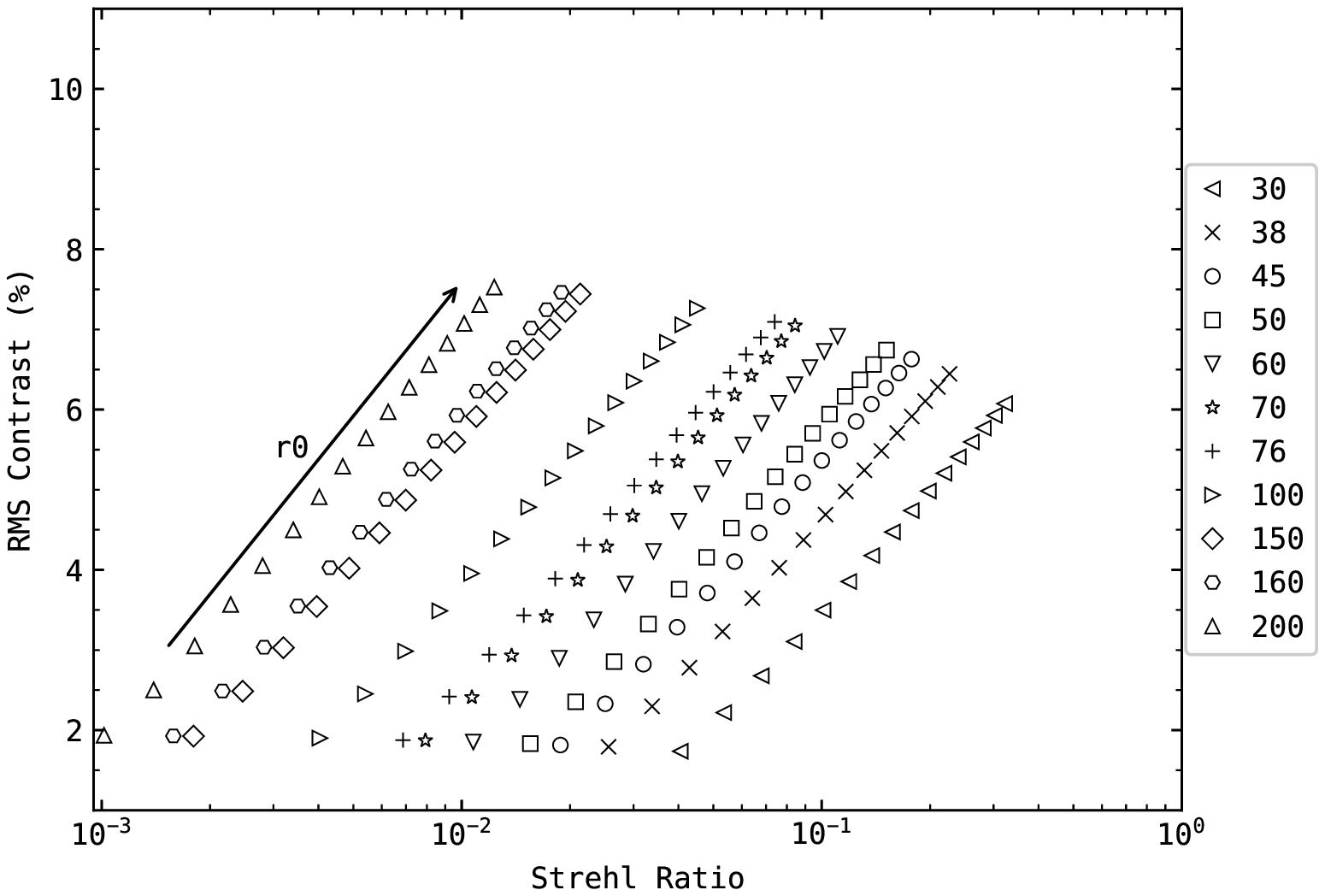}
\label{con_vs_strhl_1}
\end{figure}

Figure~\ref{con_vs_strhl_1} shows a semi-logarithmic plot of rms contrast as a function of Strehl ratio. The Strehl ratio spans over three orders of magnitude as $D/r_0$ changes from 1 to 33. However, the rms contrast changes by less than an order of magnitude for the same range of $D/r_0$. This is perhaps due to the intrinsically low contrast nature of the solar granulation. This plot helps us to specify the contrast of the solar granulation as metric for solar telescopes against the traditional Strehl ratio (which cannot be measured) for seeing-limited imaging. Conversely, it could also be used to estimate the efficiency of the telescope under seeing-limited imaging conditions (by comparing the observed contrast with the theoretical upper limit presented here).

\subsubsection{Stabilized Imaging}
\begin{figure}[H]  
    \caption{Rms contrast and Strehl ratio vs D/$r_0$ for stabilized imaging case. The markers represent different telescope diameters (in cm) as shown in the legend. }  
    \includegraphics[trim = {1.5cm, 0.5cm 1.65cm 0cm}, width=0.85\textwidth]{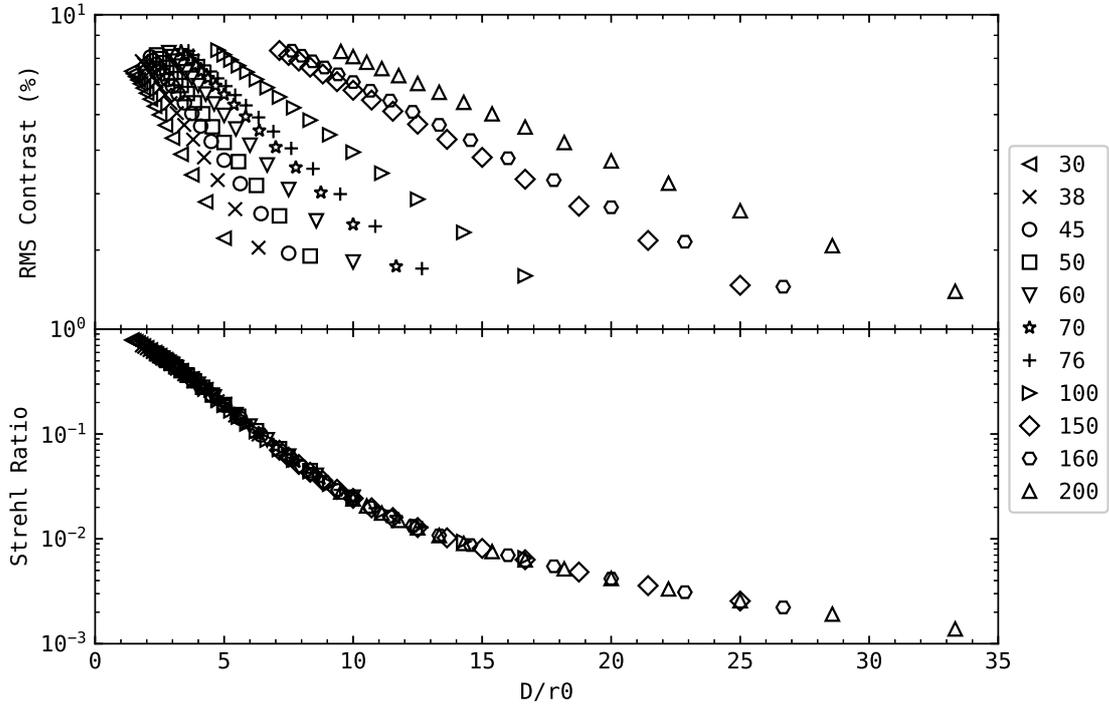}
\label{fig:post_ttc}
\end{figure}

The first order adaptive optics compensation is to stabilize the image by arresting or mitigating its random motion at kHz rate. It is equivalent to removing the 2D tilt in the atmospherically induced phase perturbations. The top and bottom panels of Figure~\ref{fig:post_ttc} indicate the rms contrast and Strehl ratio after compensating for the fast varying wavefront tilt as a function of $D/r_0$. We observe that: \begin{enumerate}[a),]

\item  The Strehl ratio is a monotonically decreasing function of $D/r_0$. It rapidly decreases when $D/r_0$$ \leq $10  but decreases relatively slowly when $D/r_0$$\ >$10. 
\item The rms contrast is a non-monotonic function. It is highly dependent on the actual telescope diameter. The rate of enhancement of rms contrast with $r_0$ (seeing) is more rapid for small and intermediate size telescopes than that for large telescopes. 
\end{enumerate}

\subsubsection{Imaging with AO correction}

\begin{figure}[H]  
    \caption{Rms contrast vs Strehl ratio for different telescope diameters and $r_0$ after AO correction. Each marker represents a unique value of N$_Z$ (shown on top left corner of the respective plots) and each colour represents a unique value of telescope diameter (shown on the right of the plots).}  
    \includegraphics[trim = {5cm 0.5cm 5cm 0.5cm}, width=0.965\textwidth]{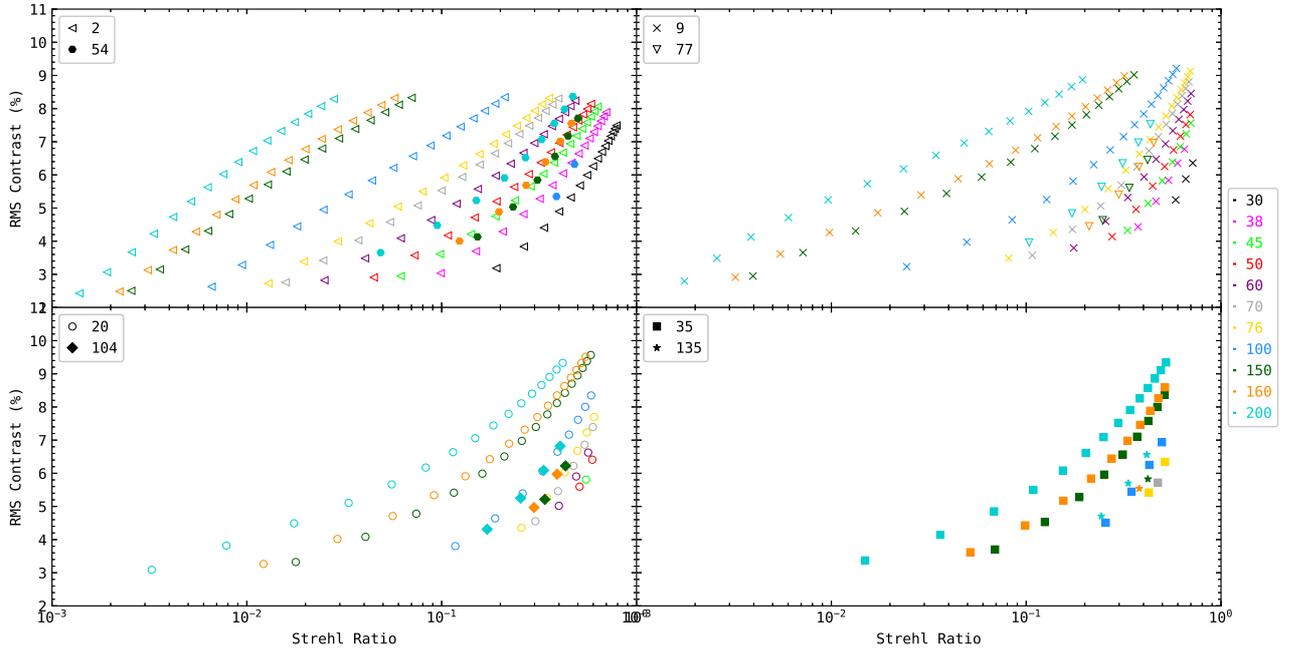}
\label{con_vs_strhl_2}
\end{figure}

The four panels of Figure~\ref{con_vs_strhl_2} show semi-logarithmic plots of rms contrast versus the Strehl ratio after AO correction. We note that the linear-log relation that existed before AO correction no longer exists. From the figure legends, we can identify N$_z$ and D as the marker shape and color respectively. For example, cyan corresponds to a 200 cm telescope and triangles and squares correspond to 2 and 35 terms corrected, respectively. $r_0$ can be found by tracing the plots along a given D and N$_Z$. For example, the cyan triangle curve in the top left panel is made up of 16 points with the lowest data point corresponding to $r_0$ of 6 cm and the highest data point corresponding to $r_0$ of 21 cm. So, for each successive point, the value of Fried's parameter increases in steps of unity. However, when we trace the cyan squares curve on the bottom right panel, we see that the maximum value of $r_0$ is only 19 cm. This is because we terminated the simulations for the cases where the phase variance was lesser than 1 radian$^2$ (Section \ref{AO_Corr}). The ``missing'' $r_0$ = 20 and 21 cm points imply that the phase variance for that $D$ and $r_0$ in the previous value of N$_Z$ (= 19) was less than 1 radian$^2$. The two red circles on the bottom left panel implies that for a 50 cm telescope, the mean square phase variance reduces to less than 1 radian$^2$ after correcting just 20 terms at $r_0$ = 7 cm; this yields corresponding Strehl and rms contrast as 0.6 and 6.5\% respectively at $r_0$ = 7 cm.  A 2 m class telescope will require compensation of up to 135 terms to bring the phase variance below 1 radian$^2$ at $r_0$ = 8 cm (blue stars on the bottom right panel). Other data points can be interpreted in a similar way.

In general, for small telescopes, both the Strehl ratio and the rms contrast increase (clustering near top right corner of the plots) with AO correction. However, for large telescopes, the increase is rather slow. Here again, this plot is quite useful to specify the granulation contrast as a metric after AO correction as against the traditional Strehl ratio.

\subsection{Discussions}

At the outset, we would like to emphasize our main result \textemdash{the semi-logarithmic plots of rms contrast of solar granulation versus the Strehl ratio.}  With this correspondence, the rms granulation contrast can be considered as the counterpart of the metrics like the Strehl ratio and encircled energy (used to specify the image quality in stellar telescopes) in solar telescopes. 

\subsubsection{Scene Dependency}
The rms granulation contrast is a function of the solar granulation scene that is being observed. At large enough fields-of-view, this variation will not be high. We find a change in the intrinsic contrast with a change in region since we are using high-resolution images covering a very small field-of-view. So, we have repeated the entire simulation for 10 different solar regions following the method described in Section \ref{simulation_flow} and obtained the mean and standard deviation of rms granulation contrast. 

\begin{figure}[H]  
    \caption{Rms contrast and Strehl ratio vs D/$r_0$ for stabilized imaging case similar to Figure \ref{fig:post_ttc} with the error bars representing the spread in values expected with change in scene.}
    \includegraphics[trim = {1.5cm, 0.5cm 1.65cm 0cm}, width=0.8\textwidth]{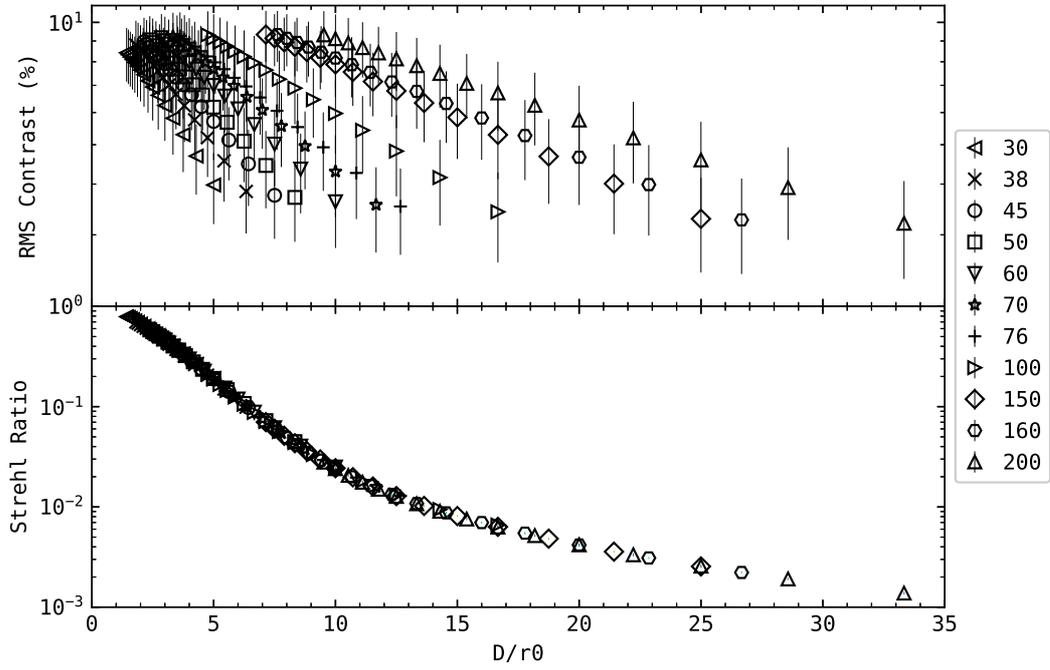}
\label{fig:post_ttc_err}
\end{figure}

\begin{figure}[H]  
    \caption{Rms contrast vs Strehl ratio similar to Figure \ref{con_vs_strhl_2} with the error bars corresponding to the deviation from mean value that can be expected when the scene of observation is changed.}
    \includegraphics[trim = {5cm 0.5cm 5cm 0.5cm}, width=0.965\textwidth]{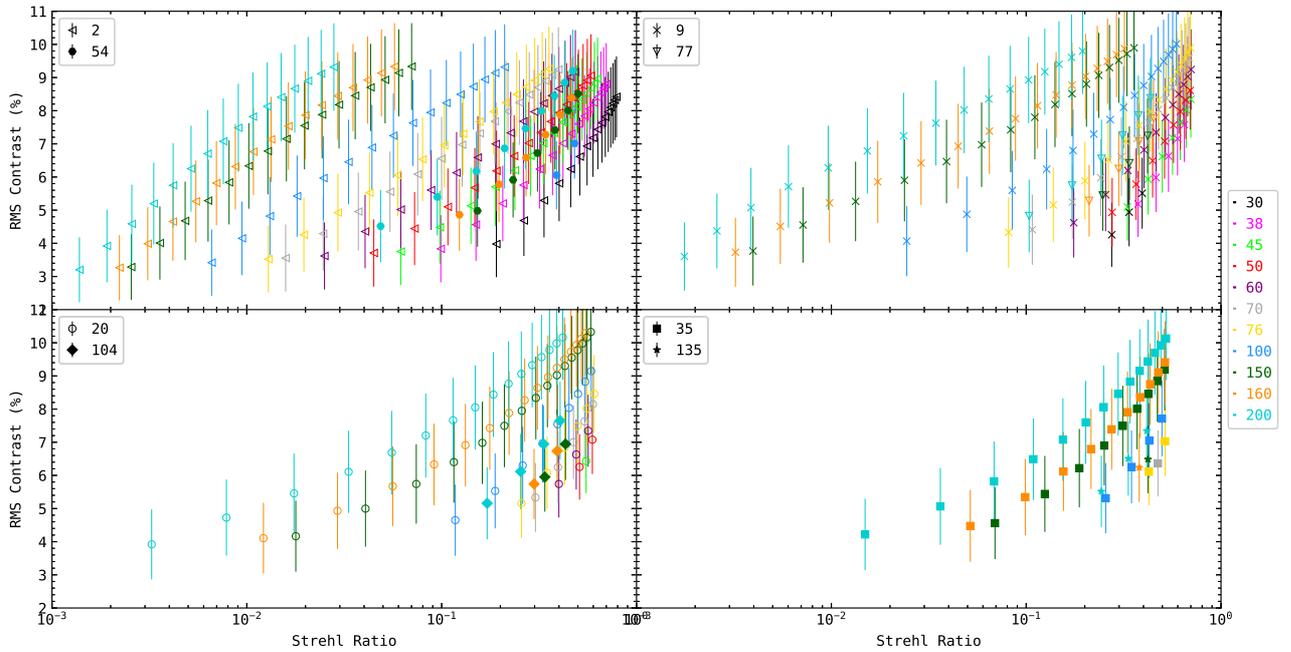}
\label{con_vs_strhl_2_err}
\end{figure}

Figure \ref{fig:post_ttc_err} is similar to Figure \ref{fig:post_ttc} showing the variation of rms contrast and Strehl ratio with D/$r_0$ for various telescope diameters. It can be seen from the top panel that the rms contrast can vary by up to 1 \% above and below the mean value. However, the Strehl is independent of scene (it depends only the transfer function of the telescope). Similarly, Figures \ref{con_vs_strhl_2} and \ref{con_vs_strhl_2_err} are comparable but for the difference due to error bars arising from scene dependency.

\subsubsection{Wavelength Dependency}
As we already know the atmospheric path-length perturbations are achromatic. However, the wavelength dependency enters our simulations through the specifications of the Fried's parameter $r_0$.  Although we have used the shortest wavelength $\lambda$ = 430.5 nm, our results for a longer wavelength can be easily obtained by changing the input $r_0$ according to its wavelength dependency of $\lambda^{1.2}$.  We have verified this through simulations as shown in Table \ref{lambda_dependancy}. It can be seen that at long wavelengths, the Fried's parameter is high and therefore the contrast is high (even though the intrinsic contrast of the granulation is lower at longer wavelength).

\begin{SCtable}[][ht]
\caption{ Comparison of the rms contrast and Strehl ratios for a 200 cm telescope when the simulations were run for two different wavelengths ($\lambda$ = 430.5 nm and 860 nm). If we consider an $r_0$ of 7 cm at 430.5 nm, this corresponds to an $r_0$ of 16 cm at 860 nm. 
}
\label{lambda_dependancy}
\begin{tabular}{cccccccc}     
    \hline                  
No. & $N_z$ &\multicolumn{2}{c}{ $\lambda$ = 430.5 nm, $r_0$ = 7 cm } &\multicolumn{2}{c}{ $\lambda$ = 860 nm, $r_0$ = 16 cm } \\
& & Rms Contrast & Strehl & Rms Contrast & Strehl \\
& & \% & Ratio & \% & Ratio \\
  \hline
1 & 0 & 2.502 & 0.0013 & 4.252 & 0.0071 \\
2 & 2 & 3.066 & 0.0019 & 5.180 & 0.0127 \\
3 & 9 & 3.486 & 0.0025 & 5.964 & 0.0826 \\
4 & 20 & 3.819 & 0.0078 & 6.673 & 0.2563 \\
5 & 35 & 4.144 & 0.0362 & 7.415 & 0.4219 \\
  \hline
\end{tabular}
\end{SCtable}

\subsubsection{Limitations of the Simulations}

The results of our simulations, particularly those with AO corrections, correspond to ideal conditions. We have ignored the finite size of the wavefront sensor and corrector elements. We have also ignored the finite temporal delay that occurs in real systems. Thus, our results are only indicative of an upper limit on the contrast and the Strehl ratios.  In what follows (Section~\ref{subsubsec:cwrs}), we compare the Strehl ratio and rms contrast obtained through our simulations with that obtained with real solar adaptive optics systems and thus derive an efficiency parameter.  We then propose to use this efficiency parameter along with the upper limits obtained through our idealistic simulations, to specify the expected Strehl ratio and hence the rms granulation contrast measured by future solar telescopes. 
A caveat in this argument is that a certain degree of efficiency in the domain of Strehl ratio need not translate to the same degree of efficiency in the rms contrast, owing to the non-linear relationship between the Strehl ratio and rms contrast after AO correction.
\cite{2010A&A...521A..68S} have reported an apparent efficiency factor of 54\% in the rms solar granulation contrast after a low order ($\approx$ 30 modes) AO correction for a 1 m telescope (factor 1.85 mentioned in Figure 5). It implies that for a larger telescope, a similar efficiency might be achieved with a high order AO correction.  It is clear that more data is required to get a better idea of if and how these parameters will affect the efficiency. Nevertheless, we can assume at least 50\% efficiency in the rms contrast and use our results as a lower bound on the contrast to be expected.

Another limitation of our simulations is that we have assumed Kolmogorov type turbulence.  In reality, the outer scale length could be finite and this would lead to a slightly better resolution \citep{2002PASP..114.1156T, 2010Msngr.141....5M}. It is also known that the residual variances after compensation of a few low order Zernike terms is lower than that predicted by the Kolmogorov turbulence even when the outer scale 10 times larger than the aperture diameter \citep{L0_paper}. Thus, real systems could be better than what is predicted based on Kolmogorov turbulence. In the same vein, metrics like the Strehl ratio and rms contrast could also be higher and better respectively. 

\subsubsection{Efficiency of Real AO Systems}
\label{subsubsec:cwrs}

We could glean the Strehl ratio obtained with three practical solar AO systems.  The first system  was that of the 70 cm Vaccum Tower Telescope (VTT) (\cite{GREGOR_AO}). Here the residual variance of the corrected modes, uncorrected modes and wavefront sensor (WFS) errors are added to determine the Strehl (Table \ref{Prac_val} - rows 1 to 6).

The second system was that the 76 cm Dunn Solar Telescope (DST). The Strehl values were estimated using three different methods. In the first method, a quasi long-exposure point spread function was estimated using the wavefront error \citep{DST_LE_PSF}. The corresponding optical transfer function was expressed as the product of three transfer functions and estimated appropriately. Finally, the Strehl ratio was estimated from the optical transfer function.

These values are listed in Table \ref{Prac_val} rows 7 and 8. The efficiency here seems to be high (compared to the German VTT case) especially for the larger $r_0$ case. One possible explanation for this is the saturation of Strehl values with an increase in $N_z$. For $r_0$ = 17 cm, it was found through our simulations that correcting for 35 terms itself will result in a Strehl of 0.86. So we opine that such extreme cases should not be considered for calculating the efficiency. 
In the second method, the Strehl (Table \ref{Prac_val} - row 10) was determined by extracting the wavefront error information after processing the AO corrected image using phase diversity method \citep{Rimmele_Sol_AO}. Here again, the apparent  high efficiency could be attributed to the combination of AO correction and image post processing and thus we exclude this case as well. Also, the values are taken from a figure in the paper. The figure only displays the best 20 out of every 100 frames. This could be another reason for high Strehl.
In the third method, the residual errors from SHWFS were used to determine the Strehl (Table \ref{Prac_val} - row 9). Here again, as stated in \cite{Rimmele_Sol_AO}, the Strehl ratios are overestimated as the contribution of higher order modes (not detected by Shack-Hartmann wavefront sensor, SHWFS) is not taken into account.

\begin{table}[H]
\caption{ Comparison of our simulated with practical Strehl values reported by solar observatories. The 70 cm VTT, 76 cm DST, and 100 cm NVST had published Strehl values of their AO systems (for method of calculation, see text). Simualtions were carried out for these telescopes with appropriate $r_0$ and $N_z$ and compared to the published values to derive an efficiency factor. Here, `Dia' refers to the telescope diameter.
}
\label{Prac_val}
\begin{tabular}{cccccccc}    

    \hline                   
No. & Dia & $N_z$ &\multicolumn{2}{c}{ Our Simulated} &\multicolumn{2}{c}{Practical} & Efficiency\\
& & & $r_0$ & Strehl & $r_0$ & Strehl& $\frac{\textrm{Our Simulated}}{\textrm{Practical}}$\\
&(cm) & & (cm) & & (cm) & &\\
  \hline
1& 70 & 27 & 9 & 0.622088 & 8.8 & 0.275 & 0.442\\
2& 70 & 27 & 11 & 0.711601 & 11 & 0.3 & 0.421 \\
3& 70 & 27 & 13 & 0.772764 & 12.5 & 0.37 & 0.478 \\
4& 70 & 27 & 15 & 0.816114  & 15 & 0.42 & 0.515 \\
5& 70 & 27 & 17 & 0.847887 & 17 & 0.45 & 0.531 \\
6& 70 & 27 & 19 & 0.871858 & 19 & 0.46 & 0.528 \\
7& 76 & 80 & 6 & 0.662532 & 5.4 & 0.46 & 0.694 \\
8& 76 & 80 & 17 & 0.92989 & 16.5 & 0.88 & 0.946 \\
9& 76 & 15 & 25 & 0.847131 & 25 & 0.8 & 0.944 \\
10& 76 & 20 & 9 & 0.497198 & 8.6 & 0.4 & 0.804 \\ 
11& 100 & 65 & 7 & 0.536508 & 7 & 0.55-0.65 & 1.025 - 1.216\\ 
12& 100 & 65 & 8 & 0.607205 & 8 & 0.6-0.7 & 0.988 - 1.153 \\
13& 100 & 65 & 9 & 0.663494 & 9 & 0.68-0.72 & 1.025 - 1.085 \\
14& 100 & 65 & 10 & 0.708692 & 10 & 0.75 & 1.058 \\
15& 100 & 65 & 11 & 0.745375 & 11 & 0.77 & 1.033 \\

  \hline
\end{tabular}
\end{table}

The third system that was considered was the AO system of the 1 m New Vaccum Solar Telescope (NVST) at the Fuxian Solar Observatory in China (\cite{Fuxian_Obs}). They  added the residual error of low-order corrected modes, high-order uncorrected modes, and aliasing error to estimate the total wavefront error. Following this, they used the expression from \citep{SR_exp} to estimate the short exposure Strehl ratio (see table \ref{Prac_val} rows 11 - 15). The values predicted by our simulations are lower than the values reported by them. Understandably, it is not a fair comparison because we estimate Strehl ratios from long exposure images.

In summary, we find that the efficiency obtained from the VTT is likely to be unbiased and thus we can possibly conclude that the efficiency of practical solar AO systems is likely to be in the range of 40 to 55\%.

\section{Summary}
\label{summary}
We have estimated, through extensive computer simulations, the rms contrast of the solar granulation to be expected from a large ground-based telescope without and with AO correction. Our simulations indicate 4\% rms granulation contrast at $r_0$ = 10 cm (at $\lambda$=430.5 nm) and 0.3\% Strehl ratio for a 2 m class telescope under seeing-limited imaging.  The rms contrast increases to 4.5\% and the Strehl to 0.4\% after image stabilization. A high order AO system with compensation equivalent to that of about 100 Zernike modes will be required to achieve a Strehl ratio of about 40\% and a rms granulation contrast of 7.5\% under similar atmospheric conditions.
We have compared our results with the existing solar AO systems and derived a possible efficiency of about 40 to 55\% for Strehl ratios.  

Although this efficiency could not be directly translated and used for obtaining observable rms granulation contrast, a similar value could be used to obtain the rms contrast to be expected from real systems in conjunction with the contrast predicted by our idealistic simulations for providing a lower bound (minimum value). Thus, our results could be quite helpful, to specify the image quality requirements for future large telescopes. 

\section*{Acknowledgements}
We would like to thank the referee for their helpful comments. 
We would also like to thank Prof S. P. Rajaguru of the Indian Institute of Astrophysics for providing the solar granulation images which were used for the simulations. This research has made use of the High Performance Computing (HPC) resources (\url{https://www.iiap.res.in/?q=facilities/computing/nova} and \url{https://www.iiap.res.in/facilities/computing/delphinus}) made available by the Computer Center of the
Indian Institute of Astrophysics, Bangalore.

\section*{Declarations}
\subsection*{Competing Interests}
The authors declare that they have no conflicts of interest.

\subsection*{Data Availability Statement}
The authors received the simulated solar granulation images from another source (Prof. S. P. Rajaguru of the Indian Institute of Astrophysics) and therefore cannot share the same. The method of generation of phase screens has been described in detail in \cite{Merak} and can be used to reproduce the phase screens.

\end{article} 

\end{document}